\documentstyle[twocolumn,prl,aps,psfig]{revtex}
\begin{document}
\twocolumn[\hsize\textwidth\columnwidth\hsize\csname@twocolumnfalse%
\endcsname
\title{Transport through Quantum Dots: Analytic Results
from Integrability}
\author{R. M. Konik$^{1,3}$,
H. Saleur$^2$, and A. W. W. Ludwig$^3$}
\address{
$^1$Department of Physics, University of Virginia, Charlottesville,
VA 22903}
\address{
$^2$Department of Physics, University of Southern California,
Los-Angeles, CA 90089-0484} 
\address{
$^3$Department of Physics, UCSB, Santa Barbara, CA 93106}

\date{\today}
\maketitle
\begin{abstract}
Recent experiments have probed quantum dots through transport
measurements in the regime
where they are described by a two lead Anderson model.  
In this paper 
we develop a new method to analytically compute for the first time
the corresponding transport properties.  This is 
done by using the exact solvability of the Anderson Hamiltonian,
together with a generalization of the Landauer-B\"uttiker approach
to integrable systems.  The latter requires proper identification
of scattering states, a complex and crucial
step in our approach.
In the Kondo regime, our results include 
the zero-field, finite temperature linear response conductance,
as well as the zero-temperature, non-equilibrium conductance
in an applied Zeeman field.

\end{abstract}
\pacs{PACS numbers: ???}  ]  
\bigskip
\vskip .4in

\newcommand{\del}{\partial}
\newcommand{\ep}{\epsilon}
\newcommand{\clsd}{c_{l\sig}^\dagger}
\newcommand{\cls}{c_{l\sig}}
\newcommand{\cesd}{c_{e\sig}^\dagger}
\newcommand{\ces}{c_{e\sig}}
\newcommand{\up}{\uparrow}
\newcommand{\down}{\downarrow}
\newcommand{\il}{\int^{\tilde{Q}}_Q d\la~}
\newcommand{\ilp}{\int^{\tilde{Q}}_Q d\la '}
\newcommand{\ik}{\int^{B}_{-D} dk~}
\newcommand{\ila}{\int d\la~}
\newcommand{\ilpa}{\int d\la '}
\newcommand{\ika}{\int dk~}
\newcommand{\tQ}{\tilde{Q}}
\newcommand{\rh}{\rho_{\rm bulk}}
\newcommand{\ri}{\rho_{\rm imp}}
\newcommand{\sh}{\sig_{\rm bulk}}
\newcommand{\si}{\sig_{\rm imp}}
\newcommand{\rph}{\rho_{p/h}}
\newcommand{\sph}{\sig_{p/h}}
\newcommand{\rp}{\rho_{p}}
\newcommand{\sip}{\sig_{p}}
\newcommand{\drph}{\delta\rho_{p/h}}
\newcommand{\dsph}{\delta\sig_{p/h}}
\newcommand{\drp}{\delta\rho_{p}}
\newcommand{\dsp}{\delta\sig_{p}}
\newcommand{\drh}{\delta\rho_{h}}
\newcommand{\dsh}{\delta\sig_{h}}
\newcommand{\enp}{\ep^+}
\newcommand{\enm}{\ep^-}
\newcommand{\enpm}{\ep^\pm}
\newcommand{\enph}{\ep^+_{\rm bulk}}
\newcommand{\enmh}{\ep^-_{\rm bulk}}
\newcommand{\enpi}{\ep^+_{\rm imp}}
\newcommand{\enmi}{\ep^-_{\rm imp}}
\newcommand{\enh}{\ep_{\rm bulk}}
\newcommand{\eni}{\ep_{\rm imp}}
\newcommand{\sig}{\sigma}
\newcommand{\la}{\lambda}

In recent years there have been a flurry of experimental studies of
quantum dots\cite{gold,cron}.
In these experiments a single localized dot state interacts with
connecting leads notwithstanding
that a {\it finite} number of electrons sits on the dot.
In a testament to advances in the miniaturization of solid state
technology, 
the occupancy of this level can be controlled by a gate voltage,
thereby controlling the current
through this `single-electron transistor'.
For an odd number of electrons, the dot becomes, due to Kramers degeneracy, 
a nanoscale realization of the single-impurity
Kondo system  -- one has in effect a `tunable' Kondo model\cite{gold,cron}.
The observed split peak structure in the non-equilibrium conductance
in an applied Zeeman field is a hallmark 
of the Kondo physics\cite{winmeir,gold,cron,ralph}.

The experimental probes of these systems are measurements of
transport properties.
The appropriate theoretical framework
to describe electronic transport through the dot is
the two-lead Anderson model 
(to distinguish it from the one-lead model appropriate for
studying dilute impurities in a bulk metal).  This model has
been examined using a variety of techniques.  In \cite{nglee},
the $T=0$ linear response conductance was studied both 
in and out of a magnetic field
via the Friedel sum rule (although the occupancy
of the dot necessary to apply the rule was computed only in
perturbation theory).  Using the NCA approximation and perturbation
theory, \cite{winmeir,hersh} studied quantum dots 
out-of-equilibrium.
However such techniques have limitations: the NCA approximation
does not work in the presence of a magnetic field while perturbation theory
requires either the dot Coulomb repulsion or
the dot-lead coupling to be small.
Numerically, the zero-field,
finite temperature linear response conductance
has been computed in \cite{costi} using the numerical 
renormalization group (NRG). 
However the latent integrability of the (one-lead) Anderson
model\cite{kao,wie} has
never been exploited to compute directly transport properties.

In this work we combine the integrability of the Anderson
model with Landauer-B\"uttiker transport theory.
To do so one must first face the challenge
of identifying scattering states in the context of integrability.
In \cite{mwen} the Bethe ansatz of the simpler one-lead
Kondo model was employed to study the equilibrium impurity DOS,
of which the non-equilibrium counterpart, in context, would
yield the desired transport properties \cite{winmeir}.
Neither the computation of the latter nor the scattering states
in the two lead model were addressed in \cite{mwen}.
The identification of the scattering
states is non-trivial.  The correct scattering states will necessarily
be electronic in nature (i.e. carry charge e and spin 1/2) and will
be confined to a single lead.  
In contrast,
the eigenstates of the 
exact solution affect a spin-charge separation (i.e. they
are not electronic) and they are not confined to a single lead.
One purpose of this letter is to outline how one can understand scattering
states in terms of the eigenstates of the Bethe ansatz.

We are successful in this endeavour on two counts.  Firstly, we are able
to construct scattering states at the Fermi surface for a dot
under
arbitrary gate voltage and arbitrary magnetic field 
and so are able to offer for the
first time a proof of the Friedel sum rule based upon
integrability, alternative to that from 
Fermi liquid theory \cite{langreth}.  
Secondly, provided
the gate voltage is adjusted so that the dot sits at the symmetric
point, we can compute scattering states away from the Fermi surface.
Through a Landauer-Bu\"ttiker approach we are then able 
to compute the zero-field finite temperature linear response
conductance (in agreement with numerical results of\cite{costi}),
as well as the $T=0$ non-equilibrium magneto-conductance, exhibiting the
observed peak splitting.

{\it Model:} To then begin, the two-lead Anderson model 
Hamiltonian is given in the continuum limit by
\begin{eqnarray}\label{ei}
{\cal H} &=&
\sum_{l\sig} \int^\infty_{-\infty} dx \{ -i\clsd (x) \del_x \cls (x)
+ V_l \delta (x) [ \clsd (x) d_\sig \cr
&& ~~~ + d^\dagger_\sig \cls (x)]\} +
\ep_d \sum_\sig n_\sig + U n_\up n_\down ,
\end{eqnarray}
where $n_\sig = d^\dagger_\sig d_\sig$.  The
$c_l$'s are the lead electrons while the $d$'s are the dot
electrons.  Here $\sum_l$ is a sum over the
two leads ($l = 1,2$).  Our formalism 
allows for the possibility that the
hopping matrix element, $V_l$, differs between the leads (as is
typical in any experimental realization).  For simplicity, in our
presented results we assume $V_1=V_2$.
Rather than treating the leads as half-lines with both left and right
moving fermions, 
we represent the leads as `unfolded' with fermions
that are solely right-moving.  
Fermions in either lead that are incident upon the dot are considered to lie
in the region, $x<0$, while those traveling away from
the dot in either lead are found at $x>0$.
We represent this in Figure 1 by drawing the leads as elongated arcs.

\begin{figure}[tbh]
\centerline{\psfig{figure=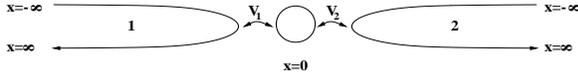,height=.35in,width=3.0in}}
\caption{A sketch of two leads attached to a quantum dot.}
\end{figure}

It will be advantageous to 
reformulate this problem as a
one-lead Anderson Hamiltonian.  To do so, we introduce
even/odd electrons:
$c_{e/o} = ( V_1 c_1 \pm V_2 c_2 )/\sqrt{V_1^2 + V_2^2}.$
Recasting ${\cal H}$ 
in this new basis, the odd electron, $c_o$, decouples and we
are left with
\begin{eqnarray}\label{eii}
{\cal H} &=& \sum_{\sig} \int dx \{ -i\cesd (x) \del_x \ces (x)
+ (V_1^2+V_2^2)^{1/2} \delta (x) \cr
&& \times \big[ \cesd (x) d_\sig + d^\dagger_\sig \ces (x)\big]\} + 
\ep_d \sum_\sig n_\sig + U n_\up n_\down .
\end{eqnarray}
We have thus reduced the problem to that studied using Bethe
ansatz in a series of papers by Kawakami and Okiji \cite{kao} and
Filyov, Wiegmann and Tsvelick \cite{wie}.

As such we summarize briefly the results of this work.  
Applying the Bethe ansatz to a system with N particles
and total spin $2S_z = (N-2M)$ yields a set
of quantization conditions describing a finite number of 
bare excitations in the system:
\begin{eqnarray}\label{eiii}
e^{ik_j L + i \delta (k_j)} &=& 
\prod^M_{\alpha = 1} { g(k_j) - \lambda_\alpha + i/2 \over  
g(k_j) - \lambda_\alpha - i/2};\cr
\prod^N_{j = 1} {\lambda_\alpha - g(k_j) + i/2 \over  
\lambda_\alpha - g(k_j) - i/2} &=& - \prod^M_{\beta=1}
{\lambda_\alpha - \lambda_\beta + i \over 
\lambda_\alpha - \lambda_\beta - i},
\end{eqnarray}
where
$\delta (k) = - 2 \tan^{-1} (\Gamma / (k-\ep_d))$,
$g(k) = (k-\ep_d-U/2)^2 / 2 U \Gamma$, and 
$\Gamma = (V_1^2 + V_2^2)$.

When $\ep_d > -U/2$ (the case $\ep_d < -U/2$ can be handled through
particle-hole transformations), the zero temperature ground
state of the system is formed from
$N-2M$ real $k_j$'s and $M$ real $\lambda_\alpha$'s.  Associated
with each $\lambda_\alpha$ is a 
pair of complex k's, $k^\alpha_\pm$, related to 
the $\lambda_\alpha$ via
$g(k^\alpha_\pm) = g(x(\lambda_\alpha) \mp iy(\lambda_\alpha))$,
$x(\lambda ) = U/2 + \ep_d - 
\sqrt{U\Gamma}(\lambda + (\lambda^2+1/4)^{1/2})^{1/2}$, and
$y(\lambda ) = \sqrt{U\Gamma}(-\lambda + (\lambda^2+1/4)^{1/2})^{1/2}$.
Roughly speaking, the real $k$'s can be thought of as charge excitations
and the $\la_\alpha$'s as spin excitations coupled to charge excitations,
i.e. bound states of electrons.  The finite temperature
ground state is considerably more complicated, involving 
an infinite hierarchy of excitations as categorized in the
`string-hypothesis' \cite{kao,wie}.

{\it Scattering States}: Under this one-lead reformulation, 
we are still able to make contact with the
scattering amplitudes of electronic excitations off the quantum dot.  Let
$T(\ep)$/$R(\ep)$ be the 
transmission/reflection amplitudes of electronic excitations
of energy $\ep$ between leads in the original two lead picture.
On the other hand, the even/odd excitations will scatter off the dot with 
some pure phase, $\delta_e(\ep )/\delta_o(\ep)$, where 
in particular $\delta_o(\ep)=0$.  
The two sets
of amplitudes are related straightforwardly:
\begin{eqnarray}\label{eiv}
e^{i\delta_e(\ep )} &=& R(\ep ) + T(\ep ) ;\cr
e^{i\delta_o(\ep )} &=& 1 =  R(\ep ) - T(\ep ) .
\end{eqnarray}
We will exploit extensively the fact that the determination of the
phase, $\delta_e$, gives $T$ and $R$ in the original problem.

To determine $\delta_e (\ep )$, we employ an energetics argument of
the sort used 
by N. Andrei in the computation of the $T=0$ magnetoresistance
of impurities in a bulk metal \cite{andrei}. 
Imagine adding an electron to the system.  
Through periodic boundary conditions,
its momentum is quantized, $p = 2\pi n / L$.  If the dot was absent,
the quantization condition would be determined solely by the conditions
in the bulk of the system and we would write, $p_{\rm bulk} = 2\pi n/L$.
Upon including the dot, this bulk momentum is shifted by a term scaling
as $1/L$.  The quantization condition is then rewritten as
\begin{equation}\label{ev}
p = 2\pi n / L = p_{\rm bulk} + \delta_e(\ep ) / L ,
\end{equation}
where $L$ is the system's length.  
The coefficient of the $1 / L$ term is 
identified with the scattering phase of the electron off the dot.

In order to determine the scattering phase of an electron (as opposed to
a spin or charge excitation), we must
must specify how to glue together a spin and a charge excitation
to form the electron.  The situation is analogous to adding
a single particle excitation in the attractive Hubbard model.
Adding a single spin $\up$ electron to the system demands that we add
a real $k$ (charge) excitation.  But  
at the same time we create a hole at
some $\la$ in the spin distribution.  The number of the
available slots in the spin distribution is determined by
the number of electrons in the system.  Adding an electron to the
system thus opens up an additional slot in the $\la$-distribution.
The electron scattering
phase off the impurity is then the difference of the 
right-moving k-impurity
momentum, $p_{\rm imp} (k)$, and the left-moving $\la$-hole 
impurity momentum
$-p_{\rm imp} (\la )$:
\begin{equation}\label{evi}
\delta^\up_e = p^\up_{\rm imp} = p_{\rm imp} (k) + p_{\rm imp}(\la ).
\end{equation}
We now turn our attention to computing these impurity
momenta.  As part of this, we will relate the impurity 
momenta to the impurity density of states (which in turn, will
allow us to prove the Friedel sum rule).  
We at the same time will compute the energy of the
$k$ and $\la$ excitations in order to parameterize scattering
in terms of energy.

In the continuum limit, these excitations are described by
smooth densities, $\rho (k)$ for the real $k$'s and $\sigma (\la )$
for the $\la$'s.  From (\ref{eiii}), equations describing these
densities can be derived in the standard fashion \cite{wie,kao}:
\begin{eqnarray}\label{evii}
\rho (k) &=& {1\over 2\pi} + {\Delta (k) \over L} + 
g'(k) \il a_1(g(k)-\la) \sig (\la); \cr
\sig (\la ) &=& - {x'(\la)\over\pi} + {\tilde{\Delta}(\la)\over L}
- \ilp a_2(\la '-\la)\sig (\la ') \cr
&& ~~~~ - \ik a_1(\la - g(k))\rho (k),
\end{eqnarray}
where $L$ is the system size and
$\Delta (k) = \del_k \delta (k)/2\pi$,
$\tilde{\Delta} (\la ) = -\del_\la 
{\rm Re}\delta (x(\la)+iy(\la))/\pi$, and
$a_n(x) = 2n/(\pi(n^2 + 4x^2))$.
$B$ and $Q$ mark out the `Fermi-surfaces' of the $k$ and $\la$
distributions.
A key observation to make of these equations is that
one can divide the densities into bulk and impurity pieces via
$\rho (k) \rightarrow \rho_{\rm bulk} (k) + \rho_{\rm imp} (k)/L$
and similarly for $\sigma (\la)$.  The impurity densities of
states contain all the information needed about degrees of freedom
living on the quantum dot.  For example the total numbers of
spin $\up$ and $\down$ electrons living on the dot 
are $n_{d\up} = \int^{\infty}_Q d\la \sig_{\rm imp}(\la )
+ \int^B_{-\infty} dk \rho_{\rm imp} (k)$ and
$n_{d\down} = \int^{\infty}_Q d\la \sig_{\rm imp}(\la )$.

The energies and momenta of these excitations can be derived
through well known techniques \cite{long}.  The energies are given by
\begin{eqnarray}\label{eviii}
\ep (k) &=& k - {H\over 2} - \int^{\infty}_Q 
d\la \ep (\la ) a_1(\la - g(k));\cr
\ep (\la ) &=& 2x(\la ) - 
\int^{\infty}_Q \ep (\la ' )a_2(\la ' - \la)\cr
&& ~~~ + \int^B_{-\infty} g'(k)\ep (k) a_1(g(k)-\la) .
\end{eqnarray}
The momenta are akin to the densities in that they divide
into bulk and impurity pieces.  The impurity momenta, the 
momenta relevant to scattering, are given by \cite{long}
\begin{eqnarray}\label{eix}
p_{\rm imp}(k) &=& \delta (k) + \il \si (\la ) 
\theta_1(g(k) - \la) ;\cr
\nonumber
p_{\rm imp}(\la )\!\! &=& \!  2 Re \delta ( x(\la)\! +\!iy(\la))
+\!\! \ilp \si (\la ') \theta_2 (\la\! -\! \la ') \cr
&&  + \ik \ri (k) \theta_1 (\la - g(k)),
\end{eqnarray}
where $\theta_n (x) = 2\tan^{-1}(x/n) -2\pi$.

We now observe a relationship between the impurity DOS and
the impurity momenta key to the actual computation 
of the scattering phases:
\begin{eqnarray}\label{ex}
\del_k p_{\rm imp}(k) &=& 2\pi \rho_{\rm imp} (k);\cr
\del_\la p_{\rm imp}(\la ) &=& -2\pi \sig_{\rm imp} (\la ).
\end{eqnarray}
The scattering phase for a spin $\up$ excitation created
from a k-particle and $\la$-hole is 
then given to be
\begin{equation}\label{exi}
\delta^\up_e = 
2\pi \int^{k}_{-\infty} dk \ri (k) 
+ 2\pi\int^{\infty}_\la d\la '\si (\la ').
\end{equation}
If $k$ and $\la$ are chosen to be at the Fermi
surface, i.e. $k=B$ and $\la=Q$, we prove the Friedel
sum rule, i.e. $\delta^\up_e = 2\pi n_{d\up}$.  Through a particle-hole
transformation, we can similarly characterize the scattering of 
spin $\down$ excitations and so also prove 
the Friedel sum rule in this case.

To discuss scattering away from the Fermi surface, reconsider
(\ref{exi}).  Suppose we want to
compute the scattering of a spin $\up$ electron with energy,
$\ep_{el}$.  If we choose the pair, $(k,\la )$, such that
$\ep (\la ) + \ep (k) = \ep_{el}$, we will have, via (\ref{exi}),
the scattering of an excitation with energy, $\ep_{el}$.
However we encounter
a problem in that there is not a unique pair, $(k,\la )$, for a given energy.
While we cannot solve the problem of choosing the correct pair
(or combination of pairs) in general, we
can make progress when
we are in the Kondo regime of the Anderson model 
(i.e. $U+2e_d \sim 0$).
In this regime we expect the scattering phase to vary on the scale
of the Kondo temperature,
$T_k$.  The electron scattering phase is determined by $\rho_{\rm imp}$
and $\sigma_{\rm imp}$, the two impurity densities.  Of the two, only
$\rho_{\rm imp}$ varies on scales on the order of $T_k$ ($\sigma_{\rm imp}$
is controlled by the much larger scale, $\sqrt{U\Gamma}$, governing
charge fluctuations).  Thus
in computing electronic scattering phases away from the Fermi surface
at zero temperature, it is natural to keep $\la = Q$, its Fermi surface
value, and vary $k$.  Specifically, to describe an electron 
with energy, $\ep_{el}$, we choose $(k,\la )$ such that
$\la = Q$ and $k$ such that $\ep (k) = \ep_{el}$.
With this ansatz, we then have restricted
the two dimensional phase space, $(\la , k)$, of potential
excitations carrying the quantum numbers of an electron to an one dimensional
subspace.  We have further arguments that suggest this ansatz
is exact at the symmetric point of the Anderson model \cite{long}.

{\it Linear Response} ($T>0$): We now try out this ansatz by examining
a quantity
that requires
us to understand scattering at finite energy:
the linear response conductance as a
function of temperature.  We compare it to Costi et al.'s
NRG results\cite{costi} and  
find excellent agreement.  This is important as
it indicates we have an essentially correct description of 
the low energy scattering
states.

Computing the linear response conductance at finite
T is a complicated matter.  We now have to compute the
scattering phases of the
glued charge and spin excitations 
in the presence
of a `thermalized ground state'; that is we must compute {\it dressed}
scattering matrices.  This ground state is no longer composed
of merely real k states and $\la$ states of bound spin and charge as 
it was at $T=0$. 
Rather all the possible solutions of the Bethe ansatz equations of the
model make an appearance.  

\begin{figure}[tbh]
\centerline{\psfig{figure=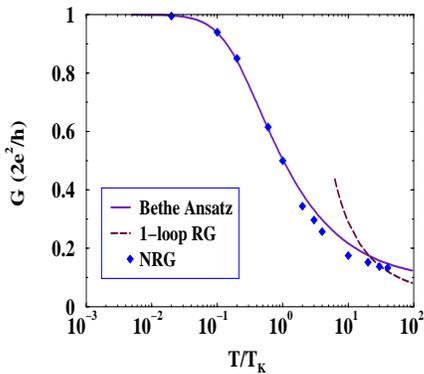,angle=-90,height=2.in,width=2.in}}
\caption{A plot of the scaling
curve for the conductance at the symmetric
point as a function of $T/T_k$.  $T_k$ is
as defined in Costi et al.'s work
and as such there are no free parameters.}
\end{figure}

However it is possible to derive equations governing
the impurity densities $\rho_{\rm imp}$ and $\sigma_{\rm imp}$
in the presence of the complicated ground state and to 
solve numerically the corresponding set
of coupled integral equations \cite{long}.
With these in hand, the linear response conductance is given by
\begin{equation}\label{exii}
G(T) \!=\! \int^\infty_{-\infty}\!\! d\ep_{el} 
(-\partial_{\ep_{el}} f(\ep_{el})) |T(\ep_{el})|^2,
\end{equation}
where $f$ is the Fermi distribution and $|T(\ep_{el})|^2=
\sin^2 ({1\over 2}\delta_{el}(\ep_{el}, T))$ is the dressed
scattering amplitude.
A plot of the result and a comparison with the NRG results of
\cite{costi} are given in Figure 2.  For temperatures
up to $T_k$, the regime where in principle an NRG computation should
be of greatest accuracy, we find excellent agreement.

Because of the Fermi liquid nature of the problem when $T\ll T_k$, 
we know the functional form of the conductance is
\begin{equation}\label{exiii}
G(T/T_k) = 1 - c T^2/{T_k}^2 + \cdots .
\end{equation}
Costi et al.\cite{costi}, based upon results borrowed from
\cite{noz,yam}, computed
$c = \pi^4 /16 = 6.088$.
We find from a fit of our curve, $c = 6.05 \pm .1$.
We have arrived at this value by fitting the plot in the region
$T/T_k < .1$.  The error is systematic in nature, arising from 
the arbitrary nature of deciding the region over which to fit.

We also compare our results in Fig. 2. with \cite{kaminski}.
It would appear that the logarithmic dependence \cite{kaminski}
characteristic of weak coupling and 
arising from a one-loop RG, should only be
expected to become qualitatively descriptive
for values of $T/T_k$ in excess of about 20.

We have recently computed the finite temperature linear response
conductance at the symmetric point in finite applied field.  
We again find reasonable agreement
with \cite{costi}.

{\it Non-Equilibrium} ($T=0,\  H \not = 0$): 
As we have successfully reproduced numerics on the finite
temperature linear response conductance, we have some confidence
that we understand scattering away from the Fermi surface.
As such we can explore
non-equilibrium conductances
which depend upon the same information.  Although there are
issues relating on how to view the 
non-equilibrium system
in the one-lead picture, these can be successfully handled
\cite{long}.
We essentially adopt the Landauer-B\"uttiker approach taken
in \cite{FLS}: we employ the in-equilibrium scattering matrices;
the sole role the bias plays is to set the particle distribution
in each lead.

\begin{figure}[tbh]
\centerline{\psfig{figure=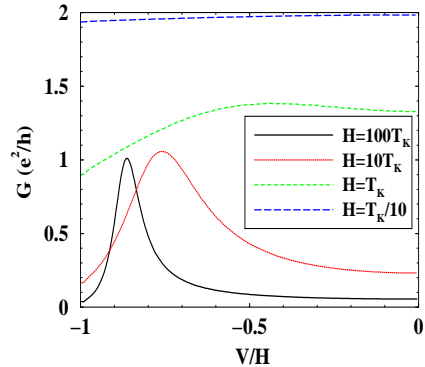,angle=-90,height=2.in,width=2.in}}
\caption{A plot of the differential
conductance in a magnetic field at the symmetric point
with $U = 8\pi\Gamma$.}
\end{figure}

The current between the two leads, lead one at zero bias and lead
two at bias $V<0$ is given by 
\begin{equation}\label{exiv}
J(V) = {e\over h}\int^{0}_{eV} d\ep 
(|T^{1\rightarrow 2}_\up(\ep)|^2 + |T_\down^{1\rightarrow 2} (\ep)|^2),
\end{equation}
In Figure 3 we plot the differential 
conductance, $G = -\del_V J$, in
the presence of a magnetic field.  Our results are in rough accordance 
with \cite{winmeir} - we find that for fields,
$H > T_k$, the 
zero-bias, zero field differential conductance peak
divides in two, one peak for each spin species.
Roughly speaking, the origin
of the split in the 
differential conductance arises from a similar bifurcation
in the impurity density of states.  The spectral weight of
the Kondo resonance present at zero energy when $H=0$ divides into
two resonances near $V \sim \pm H$, again one associated with each
spin species.  However unlike \cite{winmeir} we find that the
peak does not occur exactly at $|V| = H$ - in fact, the approach to
this value is logarithmic in $H$ \cite{long}. 
This discrepancy is  unsurprising given the
perturbative nature of the computation in \cite{winmeir}.

In the large field limit, $H\gg T_k$, our analytical 
control allows us to employ a Wiener-Hopf
analysis \cite{long} to analyze the properties of the
peak in the differential
conductance.  With $a=\log (H\sqrt{\pi e}/2^{3/2}T_k)/\pi$,
a peak maximum occurs at a bias,
$eV_{\rm max} = -H (1 - \cot^{-1}(a)/(2\pi))$.
The corresponding width of the peak is given by 
$e\Delta V = H/(2\pi) (\cot^{-1}(a-1/2)-\cot^{-1}(a+1/2))$,
while the height of peak is,
$G_{\rm max} = e^2/h(3/2-a/\sqrt{4a^2+1})$.

To conclude, 
we have shown that the integrability of the Anderson model
can be exploited to compute transport quantities
for  quantum dots.  The computation is more difficult than 
in the quantum Hall tunneling problem \cite{FLS},
and requires an ansatz
to describe scattering away from the Fermi surface. The accuracy 
obtained in reproducing known
numerical results \cite{costi} is remarkable,
and suggests that the method could be a successful way of handling
non--equilibrium transport in general integrable systems.
In future work, we will further
exploit this technique to compute 
the DC shot noise at zero temperature in quantum dot systems\cite{noise}.

R.K. has been supported by
NSERC, the NSF through grant
number DMR-9802813 and through the Waterman Award under
grant number DMR-9528578, and the University of Virginia
Physics Department.  H.S. has been supported by
the Packard Foundation, the NYI
Program, and the DOE (H.S.).
H.S. also acknowledges hospitality and support from
the LPTHE (Jussieu) and LPTMS (Orsay). AWWL has been
supported by the NSF through grant DMR-00-75064.

\end{document}